\documentclass[a4paper,10pt]{article}
\usepackage[usenames,dvipsnames]{xcolor}
\usepackage[utf8]{inputenc}
\usepackage{xspace}			   % espace intelligent dans les macros
\usepackage[english]{babel}
\usepackage[T1]{fontenc}
\usepackage{lmodern}
\usepackage{authblk}
\usepackage{enotez}

\usepackage[margin=25mm,includeheadfoot,bindingoffset=10mm]{geometry}
\usepackage{setspace}
\usepackage{ragged2e}
\usepackage[all]{xy}
\usepackage{graphicx} 

\usepackage[small,bf]{caption}

\usepackage[symbol*]{footmisc}
\usepackage{pdfpages}
\usepackage{amsmath, mathtools}
\usepackage{amssymb,amsfonts,dsfont}
\usepackage{mathrsfs}
\usepackage{amsthm}
\usepackage{bm}				
\usepackage{bbm,upgreek}
\usepackage{makecell}
\usepackage{cellspace}
\usepackage{hyperref} %backref=page
\usepackage{cleveref}

\hypersetup{%
	bookmarksopen=true,
	unicode=true,           % to show unicode characters in Acrobat’s bookmarks
	pdftoolbar=true,        % show Acrobat’s toolbar?
	pdfmenubar=true,        % show Acrobat’s menu?
	pdffitwindow=false,     % window fit to page when opened
	pdfstartview={FitH},    % fits the width of the page to the window
	pdftitle={},
	pdfauthor={},
	pdfsubject={},          % subject of the document
	pdfcreator={},          % creator of the document
	pdfproducer={pdfLaTeX}, % producer of the document
	pdfnewwindow=true,      % links in new window
	colorlinks,
	citecolor=BrickRed,
	filecolor=BrickRed,
	linkcolor=BlueViolet,
	urlcolor=BrickRed,
	linktocpage=true
}
\usepackage[sort&compress,numbers]{natbib}
\usepackage{url}
\usepackage{doi}

\newcommand{\bra}[1]{\langle #1 |} 
\newcommand{\ket}[1]{|#1 \rangle } 
\definecolor{cbl}{rgb}{0,0,1}
 
\definecolor{crd}{rgb}{1,0,0}
 
\newcommand{\upd}{\mathrm{d}}
\newcommand{\tr}{\mathrm{tr}}
\newcommand{\xb}{\mathbf{x}}

\newcommand{\ie}[0]{\textit{i.e.} }
\newcommand{\eg}[0]{\textit{e.g.} }

\newcommand\hrho{\hat{\rho}}
\newcommand\hsigma{\hat{\sigma}}
\newcommand\cGRW{\textsf{continuous}-GRW } 
\newcommand\dGRW{\textsf{deterministic}-GRW }

\title{The sound of quantum randomness}

\author{}
\date{last time compiled \small (\today) \normalsize}
\begin{document}

\noindent\LARGE \textbf{The subtle sound of quantum jumps} \normalsize
\vskip0.2cm

\vskip0.5cm
\noindent Antoine Tilloy

\noindent \textit{Max-Planck-Institut f\"ur Quantenoptik, Hans-Kopfermann-Stra{\ss}e 1, 85748 Garching, Germany} 
\vskip0.5cm

\noindent Could we hear the pop of a wave-function collapse, and if so, what would it sound like? There exist reconstructions or modifications of quantum mechanics (collapse models) where this archetypal signature of randomness exists and can in principle be witnessed. But, perhaps surprisingly, the resulting sound is disappointingly banal, indistinguishable from any other click. The problem of finding the right description of the world between two completely different classes of models -- where wave functions jump and where they do not -- is empirically undecidable. Behind this seemingly trivial observation lie deep lessons about the rigidity of quantum mechanics, the difficulty to blame unpredictability on intrinsic randomness, and more generally the physical limitations to our knowledge of reality.

\section{Introduction}

The question of whether or not the wave-function of quantum mechanics \emph{really} jumps has been posed since the early days of quantum mechanics by Schrödinger \cite{schroedinger1952}, then by Bell \cite{bell2004}, and is still feverishly debated. There is no answer to be found by staring indefinitely at the orthodox formalism and massaging its equations. It is either too vague or too cautious to say anything about reality beyond probabilistic predictions for macroscopic observations. But this does not mean that there is no way to obtain an answer at all: that quantum mechanics in its standard approach was successful \emph{despite} its metaphysical silence does not mean the latter was necessary. 

An example of an approach making simple and transparent statements about reality is the theory of de Broglie and Bohm, or Bohmian mechanics \cite{goldstein2016,durr2009}. In Bohmian mechanics, reality is made of particles which flow along a natural probability current built from the wave function. Wave-functions do not jump fundamentally, they only appear to do so effectively. Some aspects of the universe, like the initial data for the particles, will remain forever unknowable. But Nature is deterministic. That's one option.

Another option, which this story will now focus on, is that of objective (or dynamical) collapse models \cite{bassi2003}. In these models, the fundamental equation of Nature is stochastic and non-linear. The superposition principle, embodied in the unitarity of quantum mechanics, is explicitly broken. Wave-functions collapse for real, and the collapse is intrinsically random (or at least left without a more fundamental explanation at the moment). Collapse models apparently go beyond reconstructing quantum mechanics; they modify it, they bend it, even at the empirical level.

For collapse models, it thus makes sense to reverse the question of Schrödinger. Indeed, the answer to the original question is in a way too obvious: certainly and by construction, the wave-function jumps. That's settled. But could these jumps be witnessed? If quantum mechanics is modified to make wave-functions jump at the fundamental level, what trace of it, if any, can we see empirically? More subtly, if we can see or hear something, can we know for sure what caused it? Perhaps counter-intuitively, this last question is trickier. There is no possible smoking gun, no characteristic signature of these objective wave-function pops that distinguishes them from the blips of world splittings. There are simple limits to what we can know about the laws of nature. 

\section{Making collapse objective}
Let us first understand what collapse models are and what they aim to achieve. Collapse models come in many flavors and there is now a whole zoo of them \cite{bassi2013review}. However, they all take the form of a small non-linear and stochastic modification of the Schrödinger equation
\begin{equation}\label{eq:generalcollapse}
    \frac{\upd}{\upd t} \psi_t = - \frac{i}{\hbar} H \, \psi_t + \varepsilon(\psi)\; ,
\end{equation}
where $H$ is what the system Hamiltonian would be without unorthodox modification (\eg the Standard Model Hamiltonian or one of its low energy approximations), and $\varepsilon$ is small in the sense that it drives the dynamics on timescales that are far longer than typical microscopic ones.

\begin{figure}
    \centering
    \includegraphics[width=0.99\textwidth]{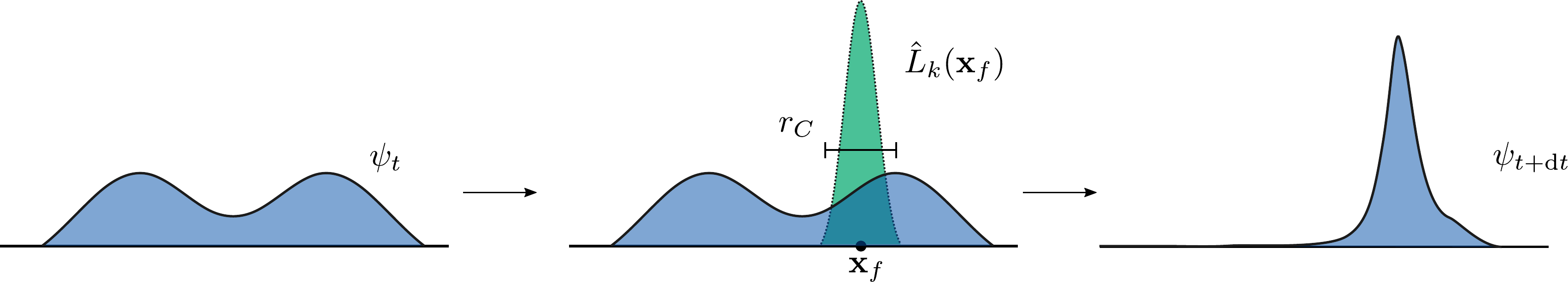}
    \caption{Random collapse induced by the GRW dynamics (shown for a single particle for simplicity).}
    \label{fig:GRW}
\end{figure}

A beautifully simple instance of this idea was put forward by Ghirardi, Rimini, and Weber (GRW) in 1986 \cite{ghirardi1986}. In the GRW model, the standard unitary evolution of orthodox quantum mechanics is interrupted by spontaneous jumps of the wave-function. Consider a $N$-particle system described by a wave function $\psi_t(\xb_1,\dots,\xb_N)$.  In the GRW model, the Schr\"odinger evolution $\partial_t \psi_t = -i(H/\hbar) \psi_t$ is interrupted by spontaneous jumps of the wave-function that can hit each particle independently. Mathematically, the jump multiplies the wavefunction in the following way:
\begin{equation}\label{eq:jump}
    \psi_t \longrightarrow \frac{\hat{L}_k(\xb_f) \psi_t}{\|\hat{L}_k(\xb_f) \psi_t\|} \, ,
\end{equation}
where the ``jump operator'' associated to particle $k$ is simply a Gaussian envelope centered around the collapse point $\xb_f$ (see Fig. \ref{fig:GRW}):
\begin{equation}
\hat{L}_k(\xb_f)=\frac{1}{(\pi r^2_C)^{3/4}} \mathrm{e}^{-(\hat{\xb}_k-\xb_f)^2/(2 r^2_C)} \, .
\end{equation}
These $N$ types of jumps occur independently of each other and are uniformly distributed in time with intensity $\lambda$.  Finally, if a jump hits particle $k$ at time $t$, the probability density for the center of the collapse $\xb_f$ is given by $P(\xb_f)=\|\hat{L}_k(\xb_f) \psi_t\|^2$, which will later prove fundamental to ensure the consistency of the model.

The width $r_C$ of the collapsed wavefunction and the frequency $\lambda$ of the jumps are two new parameters that, once empirically fixed, would fully specify the model. They have to be such that collapse events go unnoticed for small $N$, making the unitary evolution approximately valid, but large enough that macroscopic superpositions with large $N$ are almost immediately collapsed. The standard  (partially arbitrary) values proposed in the literature, $\lambda=10^{-16}\mathrm{s}^{-1}$ and $r_C=10^{-7}\mathrm{m}$, ensure that this is the case \cite{ghirardi1986,feldmann2012}, but the range allowed is narrowing quickly.

Knowing what the GRW model says about reality requires an additional choice. Now that measurement results are no longer primitive, one needs to decide what is, \ie say what the world is made of, \ie pick an ontology \cite{allori2008,allori2015} (or beables in the language of Bell \cite{bell1976}). It can be the wavefunction itself $\psi_t$ (\textit{a.k.a.} $\text{GRW}_0$, like the Many-World interpretation but with only one world), the quantum average of the mass density ($\text{GRW}_m$), or the collapse events $(\xb_f,t_f)$ ($\text{GRW}_f$). I have a preference for the last choice, because it provides a neat basis for special relativistic \cite{tumulka2006,tumulka2020} and gravitational \cite{tilloy2018} generalizations, but also, I confess, because I enjoy the minimalism of a reality made of $0$-dimensional space-time objects (instead of $d=1$ particle trajectories or $d=2$ string tubes). This sometimes feels like a purely aesthetic choice. While it is indeed empirically undecidable, we will realize that it is just as undecidable as the very choice of stochastic equation to start with.

What does the model imply about the way we do physics and make predictions in practice? First, the measurement postulate is no longer needed. Qualitatively, the model explains why the wave function of macroscopic objects, among which are measurement devices, collapses into definite positions, and thus how we obtain definite outcomes. Quantitatively, one recovers the Born rule. In practice, one can still use the standard formalism (with an effective measurement induced collapse) but with a slight modification (non-linear and stochastic) of the equation of motion for the quantum system under study.

\section{The stiff price of consistency}
An unappealing feature of collapse models, at least at first sight, is the tremendous amount of freedom there seems to be in modifying the Schr\"odinger equation with non-linear terms as in~\eqref{eq:generalcollapse}. Ghirardi, Rimini, and Weber picked one, but could they have just picked anything, something deterministic for example? In 1989, Steven Weinberg essentially thought so and imagined a class of deterministic modifications of quantum mechanics \cite{weinberg1989_annals,weinberg1989}. Alas, as Nicolas Gisin had discovered earlier the same year~\cite{gisin1989}, consistency comes at a price. While GRW had paid it, Weinberg had not~\cite{gisin1990}.

Carelessly modifying the Schr\"odinger equation indeed tends to shatter the whole probabilistic apparatus of quantum theory. In quantum mechanics, linearity makes sure that proper mixtures (probabilistic ensembles of pure states) are indistinguishable from improper mixtures (mixed states, obtained by tracing out some degrees of freedom). Why should we care about this technical indistinguishability? Among other things, it ensures that if Alice and Bob share an entangled pair of photons, the non-local correlations that manifest themselves in the violation of Bell's inequality stay benign, miraculously compatible with special relativity. Indeed, in standard quantum mechanics, Bell non-locality does not allow faster than light signalling; entanglement cannot be used as a force by Alice to send energy or information on Bob's side. This all breaks down with a deterministic non-linearity, which lifts the veil of ignorance between the two. If Alice measures the photon on her side, she changes Bob's state from an improper to a proper mixture which now has different dynamics. Whether Alice measures or not thus has an influence on Bob's side and leaves a clear trace that Bob can see \cite{bassi2015}. Entanglement no longer just creates surprising correlations but becomes a force, independent of distance, that allows to send information.

I should say that this does not fully close the door to reckless non-linear modifications of the Schr\"odinger equation. One can pray that the Born rule is also modified in precisely the way that would avoid signalling. Without precise examples, this is nothing but wishful thinking. Further, without a simple probabilistic rule, extracting predictions from a theory far more massively non-local than quantum mechanics becomes a daunting if not impossible task. Gisin's no-go for non-linear modifications is powerful, and although not fully airtight (no-go theorems rarely are), it has not been convincingly bypassed yet.

How does the GRW model keep the non-local demon in the box? In an experiment, one can measure probabilities of certain outcomes $\pi_k= \bra{\psi} \hat{\Pi}_k \ket{\psi}$ . This does not change with collapse models which, as we argued, allow to derive the standard formalism of quantum mechanics by applying the new dynamics to measurement devices. The novelty with GRW is the random term in the evolution. The crucial observation is that the probabilities of outcomes we can measure are necessarily averaged over this randomness. Indeed, we do not have an \emph{a priori} knowledge of the underlying stochastic process, and thus we have empirical access only to $\bar{\pi}_k = \mathbb{E}[\pi_k]$, where $\mathbb{E}[\, \cdot \,]$ denotes the average (in the standard statistical sense) over the GRW randomness. We can rewrite this probability to introduce the density matrix $\hat{\rho} := \mathbb{E}\left[\, \ket{\psi}  \bra{\psi}\, \right]$ :
\begin{equation}
    \bar{\pi}_k=\mathbb{E}\left[\bra{\psi} \hat{\Pi}_k \ket{\psi}\right]=\tr \left(\hat{\Pi}_k\,\mathbb{E}\big[ \ket{\psi}\bra{\psi}\big]\right) = \tr \left( \hat{\rho}\,  \hat{\Pi}_k\right).
\end{equation}
Hence, the density matrix encodes \emph{every single} empirical prediction we can make, all the probabilities of all the results of all the possible experiments we can make. \emph{Everything}\endnote{Lajos Di\'osi has been making this point regularly, likely since before I was born. See \eg \cite{diosi2017} for a recent summary of his views.}. Thus, provided the density matrix evolves linearly, we are good with Gisin's argument. In fact, using the jump equations \eqref{eq:jump} and associated jump probabilities, it is easy to see that for the GRW model, non-linearities ``miraculously'' cancel and the matrix obeys a simple \emph{linear} equation:
\begin{align}
\frac{\upd}{\upd t}  \hat{\rho}_t &= -\frac{i}{\hbar}[\hat{H},\hat{\rho}_t] + \lambda \sum_{k=1}^N \left\{ \int  \upd \xb_f  \hat{L}_k(\xb_f) \hat{\rho}_t \hat{L}_k(\xb_f) \right\}  - \hat{\rho}_t \, , \label{eq:master} \\
&\equiv \mathcal{L}\hat{\rho}_t\; , 
\end{align}
where $\mathcal{L}$ is a linear operator of the Linblad form. All other consistent (Markovian) collapse models obey a linear equation at the density matrix level. In the end, adding objective jumps into the Schr\"odinger equation can be made consistent. The price is stiff: non-linearity is allowed at the wave-function level, but it has to vanish exactly at the density matrix level.

%More generally, the Born rule means that certain objects in a theory encode probabilities. Since probabilities are necessarily linear for consistency, these objects in the theory cannot evolve non-linearly.

\section{Linearity buries the unraveled}

We started with a model with jumps, non-linearly collapsing the wave-function. In such a universe, randomness and discontinuous behavior undoubtedly lie at the deepest level of reality. But can any of these fundamental features be empirically observed? Can I see a trace of the jumps in an experiment that would convince me that they exist and come from the model?

As we just learned, every quantitative prediction one can ever possibly make with the GRW model is in its density matrix $\hat{\rho}$ which obeys the linear equation~\eqref{eq:master}. So the question becomes: what trace of the jumps remains in this equation? The answer is a disappointing one: nothing \emph{specific} at all. The linearity of \eqref{eq:master} makes the existence of the jumps empirically undecidable.

This comes from a result well known in quantum optics: there are infinitely many different ways to \emph{unravel} a Lindblad equation $\partial_t \hat{\rho}_t =\mathcal{L} \hat{\rho}_t$. In this context, an unraveling is a stochastic Schr\"odinger equation for a quantum state $\ket{\tilde{\psi}}$ such that, upon averaging $\mathbb{E}\left[\, \ket{\tilde{\psi}}  \bra{\tilde{\psi}}\, \right]$, we recover the solution of Lindblad equation. 

The GRW Lindblad equation \eqref{eq:master} is no exception, and the jump equation \eqref{eq:jump} provides \emph{one} of its unravelings. But there exist wildly different ones as well. We can give an example that is in fact both unitary (so without collapse) and smooth:
\begin{equation}\label{eq:unitaryunraveling}
    \frac{\upd}{\upd t} \psi_t = - \frac{i}{\hbar} H\,\psi_t + i\, \sqrt{\lambda} \int  \upd \xb_f \, w^k_t(\xb_f)\, \hat{L}_k(\xb_f) \,\psi_t,
\end{equation}
where $w^k_t(\xb)$ are independent white noise processes in space and time. It is easy to show\endnote{The equation \eqref{eq:unitaryunraveling} is understood in the Stratonovich convention, which corresponds to the intuitive limit of taking a noise with a correlation time much smaller than other dynamical timescales. To derive the master equation \eqref{eq:master}, one just needs to go to the It\^o representation and use It\^o's lemma to differentiate the product $\ket{\psi}\bra{\psi}$. In the It\^o representation, terms proportional to the noise are $0$ on average and thus disappear when taking the statistical average, yielding \eqref{eq:master}.} that this stochastic Schr\"odinger equation indeed yields the same master equation \eqref{eq:master} as the GRW jump prescription \eqref{eq:jump}. Let us call this new dynamics \cGRW.

What would \cGRW tell about the world if it were fundamental? It would say that the quantum state $\psi_t$ evolves \emph{unitarily} but is subject to a noisy potential (random in space and time) yielding a continuous evolution. Clearly, this is wildly different from what the GRW model says. Clearly also, \cGRW would not solve the measurement problem, and one would presumably have to understand the emergence of classicality in a Many-World way. But, empirically, \cGRW gives the exact same predictions as GRW.

In fact, we can also reproduce the predictions of GRW in another way with a (slightly contrived) deterministic and fully quantum theory. The idea is to exploit the fact that a truly random noise is always, at least empirically, indistinguishable from a coupling to a Markovian bosonic environment, or ``quantum noise''. Let us make our own makeshift quantum noise theory rapidly. The details do not really matter, the only thing that matters is to understand that it is possible. The idea is to introduce a continuum of quantum harmonic oscillators (indexed by a frequency 
$\omega$) in every-point of space $\xb$, and associate a different continuum to each particle $k$. We can introduce the state $\ket{0}$ where all oscillators have zero excitation and the associated creation and annihilation operators $a^{\dagger}_k(\omega,\xb),a^{~}_{k'}(\omega',\xb')$ which verify the canonical commutation relations
\begin{equation}
    [a^{~}_k(\omega,\xb),a^\dagger_{k'}(\omega',\xb')] = \delta^3(\xb-\xb')\delta(\omega- \omega')\delta_{k,k'} \; .
\end{equation}
Now, if we consider an extended Hilbert space $\mathscr{H}_\text{tot}=\mathscr{H}_\text{particles}\otimes\mathscr{H}_\text{oscillators}$ containing our initial $N$ particles and these oscillators, start from a state $\ket{\Psi}=\ket{\psi}\otimes\ket{0}$ at the origin of time, and have the whole thing evolve with $H_\text{tot}=H + H_\text{int} + H_\text{bath}$ with
\begin{align}
    H_\text{bath} &= \int_{\mathbb{R}^4} \upd \omega \, \upd\xb\,    \hbar \,\omega \, a^\dagger_k(\omega, \xb) a^{~}_k (\omega,\xb)\, , \label{eq:bath1} \\
    H_\text{int}&=\int_{\mathbb{R}^4} \upd \omega \, \upd\xb \, \hbar\,\sqrt{\lambda} \, \hat{L}_k(\xb) \otimes \left[a^\dagger_k(\omega,\xb) + a^{~}_k(\omega,\xb)\right] \, , \label{eq:bath2}
\end{align}
we get a world that is empirically indistinguishable from GRW. More precisely, the density matrix $\rho_\text{particles} = \tr_\text{bath}\left[\ket{\Psi}\bra{\Psi}\right]$ is easily shown to obey the GRW Lindblad equation \eqref{eq:master}.  This other version, let us call it \textsf{deterministic}-GRW cannot be told apart from GRW.

What would \textsf{deterministic}-GRW tell about the world if it were fundamental? It would tell us that Nature is still perfectly well described by quantum mechanics, without any breakdown of the superposition principle. The departure would be not so much from quantum theory in general as from the Standard Model of particle physics, with the addition of a peculiar bosonic dark sector. 

\begin{figure}
    \centering
    \includegraphics[width=0.99\textwidth]{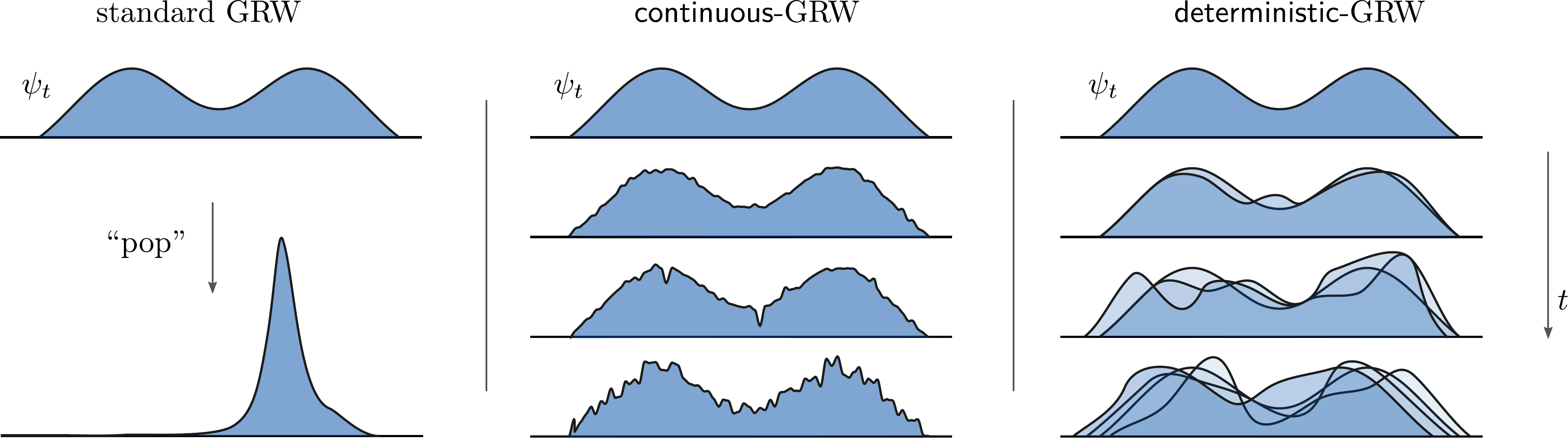}
    \caption{Comparison between the standard GRW model and its two cousins, the \cGRW and \dGRW models. For the latter, the wavefunction gets entangled with a bosonic environment, and thus becomes a superposition $\ket{\Psi} = \sum_k \ket{\psi_k}\otimes\ket{\text{env}_k}$.}
    \label{fig:cousins}
\end{figure}

Let us summarize. The predictions of the GRW model, where wave-functions undergo random discrete jumps, can be exactly reproduced by other models that tell a different story as shown in Fig. \ref{fig:cousins}. In the first, \cGRW, the evolution is still random but continuous and unitary, hence without any collapse of macroscopic superpositions. In the second, \textsf{deterministic}-GRW, the evolution is no longer random, quantum mechanics does not break down at all, only the particle content is modified. That such different stories are possible does not mean the GRW model is not falsifiable, at least for a fixed choice of parameters. Rather, the existence of simple alternative stories shows that, in the unlikely event that we observe deviations from the Standard Model that are precisely of the kind suggested by the GRW master equation, it does not follow that nature has to be random and discontinuous. In fact, it is even likely that the first intuition of physicists would be to look for a dark sector to explain the effects observed. In the case of the GRW master equation, this dark sector is sufficiently bizarre that we might prefer the original GRW dynamics, but the unitary white-noise version we discussed first is trickier to dismiss on such ``no-awkwardness'' arguments\endnote{For non-Markovian collapse models, the corresponding oscillator dark sector becomes more physical, and the deterministic unitary story would likely always be favored.}. The choice of the GRW model instead of its cousins is as metaphysical as the choice between $\text{GRW}_0$, $\text{GRW}_m$, or $\text{GRW}_f$.  This underdetermination of the ontology of a theory from its empirical content is no surprise in general, but rarely are the descriptions so brutally different and independently compelling.

\section{The sound of collapse}

To show how surprising it is to have such different models with an identical empirical content, it is instructive to look at a regime of parameters considered by Feldman and Tumulka \cite{feldmann2012}. Imagine that the collapse radius $r_C$ in the GRW model is extremely tiny, say less than $10^{-16}$m. Then each collapse would release so much kinetic energy that heat and even sound would be instantly emitted. Such a small value of $r_C$ is already falsified by other means and thus the emission of sound, even faint, is excluded empirically. But let us imagine for an instant that we live in a universe in which it is not, and where each random collapse injects enough energy to make an audible bang.

In this universe, the standard GRW story goes, we will hear random bangs, distributed in a Poissonian way, and nothing in between, or at least nothing unusual. But what would happen with \cGRW where there are no jumps and only white noise? Would we hear a faint but constant buzzing? And with \textsf{deterministic}-GRW, what would we even hear, a more complicated and structured sound, or nothing at all? But wait, that makes no sense. Our intuition deceives us. While different ``unravelings'' can yield different metaphysics, they have to agree empirically! The stories for each model have to be strictly identical if we trust the previous discussion (and we should). So what is it? Do we hear a bang, a faint buzzing, or something else?

\section{A repeated interaction apart\'e}

Let us first definitely convince ourselves that the sounds should be identical. To this end, it helps to take a step back and consider a simpler situation of \emph{repeated interactions}.

\begin{figure}
    \centering
    \includegraphics[width=0.99\textwidth]{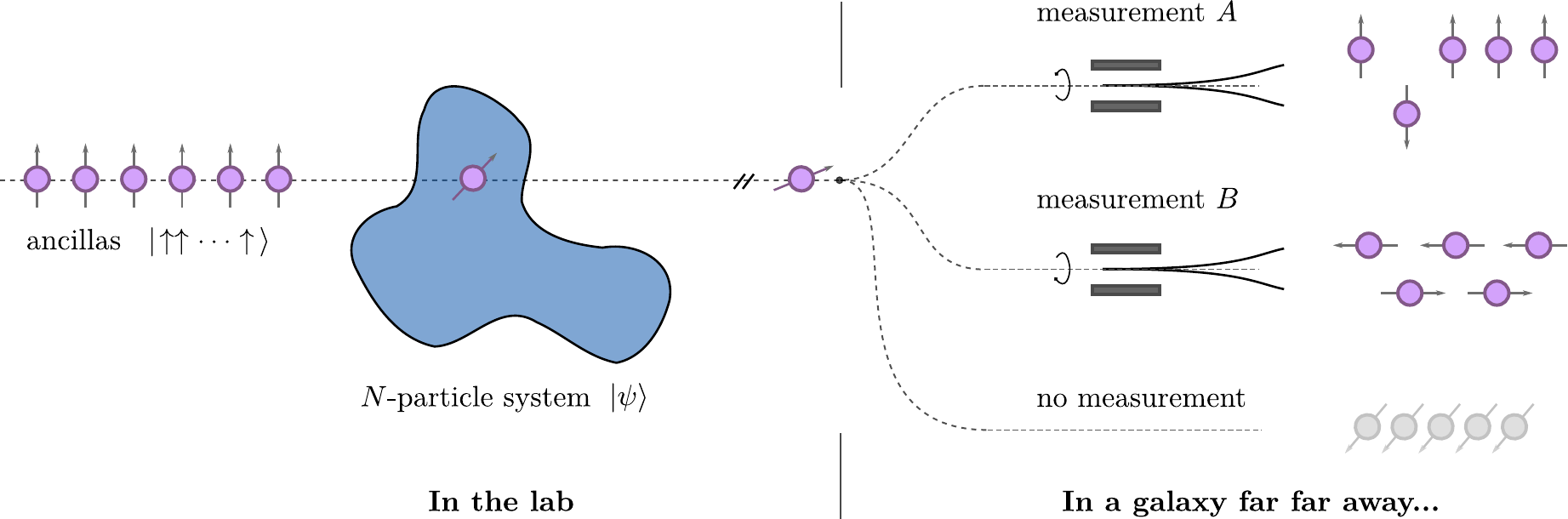}
    \caption{A simple repeated interaction setup -- Ancillas are sent one by one to a system of interest, get entangled with it, and then are sent very far away, \eg in another galaxy. There, they can be subjected to different measurements or left untouched. }
    \label{fig:repeated_interactions}
\end{figure}

Repeated interactions are a way to model Markovian open quantum systems in a discrete time setting. The idea is to have ancillas (here to simplify qubits) interact for a brief amount of time, and only once, with a system of interest (here the $N$ particles we wish to model). More precisely, we start with a bunch of two-level systems in a product state $\ket{\!\uparrow \uparrow\cdots \uparrow \,}$, and have them interact unitarily with our $N$ particle state one by one:
\begin{equation}
    \ket{\psi}\otimes\ket{\!\uparrow\,}\overset{U}{\longrightarrow} \; U_\uparrow\, \ket{\psi} \otimes \ket{\!\uparrow\,} + U_\downarrow\, \ket{\psi} \otimes \ket{\!\downarrow\,}.
\end{equation}
Unitarity implies only that $ U^{~}_\uparrow U_\uparrow^\dagger + U^{~}_\downarrow U_\downarrow^\dagger = \mathds{1} $. The setup is illustrated in Fig. \ref{fig:repeated_interactions}. The first ancilla and our state of interest are now entangled. Far away from the interaction region, we can decide what to do with the ancillas. For example we can choose to measure the ancillas in the $\{ \ket{\!\uparrow\,},  \ket{\!\downarrow\,} \}$ basis, which would yield:
\begin{equation}
    \ket{\psi}\rightarrow \left\{\begin{array}{lll}
        \frac{U_\uparrow\, \ket{\psi}}{\sqrt{\bra{\psi} U_\uparrow^\dagger U_\uparrow^{~}\ket{\psi}}} &  \text{with probability} & p_\uparrow = \bra{\psi} U_\uparrow^\dagger U_\uparrow^{~}\ket{\psi} \\
          \frac{U_\downarrow\, \ket{\psi}}{\sqrt{\bra{\psi} U_\downarrow^\dagger U_\downarrow^{~}\ket{\psi}}} &  \text{with probability} & p_\downarrow = \bra{\psi} U_\downarrow^\dagger U_\downarrow^{~}\ket{\psi}
    \end{array}\right. \, .
\end{equation}
After each measurement, ancilla and system are again in a product state, and we have thus a discrete stochastic quantum trajectory for the system $n\rightarrow \ket{\psi_n}$. Other choices of measurement of the ancilla would yield different stochastic trajectories.

Let us now imagine that the ancillas are simply never measured and that we never have access to them (\eg they are sent to another galaxy). Then all we can know about our system is in its density matrix $\hat{\rho}:=\tr_\text{ancillas}[\,\ket{\Psi}\bra{\Psi}\,]$, where $\ket{\Psi}$ is the full system + ancillas state. The latter evolves at each step simply as:
\begin{equation}\label{eq:cptp}
    \rho_{n+1}:= \Phi(\hrho) = U^{~}_\uparrow \hrho_n U_\uparrow^\dagger + U^{~}_\downarrow \hrho_n  U_\downarrow^\dagger \, ,
\end{equation}
where $\Phi$ is a completely positive trace preserving map, \textit{a.k.a.} a quantum channel, \textit{a.k.a.} a discrete version of a Linblad evolution. We could also assume that once sent in this other galaxy, the ancillas are measured in a certain basis, but that the results remain unknown to us. Then, all we could ever predict is again in the density matrix defined as $\hsigma:= \mathbb{E}[\,\ket{\psi}\bra{\psi}\,]$. As the reader now certainly expects, $\hsigma$ verifies the same equation \eqref{eq:cptp} as $\hrho$. Naturally, without access to the ancillas, what happens far away (measurement in a basis or another, or no measurement at all) does not matter. In fact, we could even choose which measurement to make on the ancilla billions of years after their interaction with the system.

This repeated interaction setup is discrete, but by making the interactions weaker and weaker (\ie $U$ close to $\mathds{1}$) and more and more frequent, it is possible to rigorously derive a continuum limit \cite{Attal2006}. We would then get a Lindblad evolution for the density matrix obtained by tracing away the ancillas (or equivalently averaging over measurement results). Different measurement schemes on the ancillas would yield different stochastic trajectories, which in the continuum are simply different unravelings of the same master equation. While the discrete setup is ``jumpy'' by definition for the stochastic system state, the continuum limit can yield jumpy or continuous dynamics depending on the choice of basis in which the ancillas are measured. 

In fact, it is possible to reconstruct all Markovian collapse models this way, as the continuum limit of a repeated interaction setup. The GRW and \cGRW models are simply associated to different measurements made on the ancillas, while the \textsf{deterministic}-GRW model is obtained by keeping the ancillas as they are\endnote{The peculiar bath defined in equations \eqref{eq:bath1} and \eqref{eq:bath2} is chosen in such a way that bosonic modes in time (Fourier transform of the frequency) interact once only with the system, effectively emulating fast repeated interactions with ancillas.}. Naturally, in this context, what is done to the ancillas becomes a fundamental law of nature, not something done by observers.

\section{Painting or carving reality into the wavefunction?}

Now that we are convinced the models are empirically equivalent, what do we hear? We can get convinced that we would hear a bang, as the standard GRW intuition suggests. What happens for the sneaky cousins?

To simplify, there are two situations: when the bang has not yet been heard and when it has. These are macroscopically distinct situations, \ie they correspond to wave functions with supports very far apart in configuration space. This is intuitive: many things are moved by the bang, if only the eardrums of those who listen. In the GRW model, a collapse makes us jump from one region to the other instantly: we hear a punctual bang. In the cousins, the weight in the wavefunction bleeds progressively from the ``no bang'' to the ``bang happened'' region. This means we have a Schr\"odinger cat state, which includes observers themselves. Since these two regions of configuration space are so widely apart, the standard decoherence story predicts a punctual bang as well for the GRW cousins\endnote{In fact, the Continuous Spontaneous Localization model, which as its name suggests is clearly continuous, is known to predict the emission of X-Ray photons from germanium \cite{piscicchia2017}. This is a finer test than the spontaneous emission of sound, but in the end it is the same idea. These X-Ray photons would be emitted punctually, showing again that the perception that something is jumpy need not be related to an underlying jump process.}.  The three models sound the same essentially because a bang from an intrinsically random jump process is empirically indistinguishable from the blip of a wave-function continuously splitting into macroscopically distinct branches.

\begin{figure}
    \centering
    \includegraphics[width=0.99\textwidth]{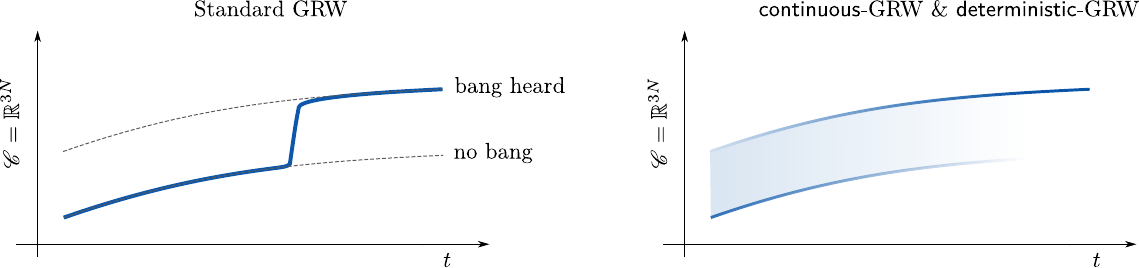}
    \caption{Illustration of the support of the wavefunction in configuration space in the standard GRW model and its cousins. All the models carve the wavefunction into two main branches, far apart, but only the GRW model paints a single trajectory.}
    \label{fig:carving}
\end{figure}

The GRW model and its variants \emph{carve} the wavefunction in the same way, digging the two macroscopically distinct regions and transferring the weight progressively from one to the other.  The difference is that the standard GRW model also \emph{paints} the branch a single world follows (see Fig. \ref{fig:carving}). While the painting is ontologically fundamental, it is unfortunately irrelevant for all practical purposes. In the repeated interaction picture, the carving corresponds to the interactions of the ancillas with the system, while different paintings are associated to the measurements made far away.

Now, let us assume the GRW model is correct. This forces reasonable values of its parameters which implies that no sound can be heard, and more generally the influence of the collapse process is far weaker than that of other forces. In practice, the macroscopic branches of the universe wavefunction are almost exclusively carved by electromagnetism, with the collapse process merely scratching the surface of the ridges and valleys. Sure, the collapse model paints the branch chosen, which is certainly ontologically useful, and makes the story -- the way we connect mathematics with reality -- far simpler than in the Many-Worlds approach. But the collapse process contributes almost nothing to the emergence of classicality for all practical purposes. Even if collapse models are correct, they are not the reason we do not see or experience macroscopic superpositions. They are the reason these macroscopic superpositions do not \emph{exist} at all.

The preferred role position seems to have in defining macroscopicity is likely also entirely explainable by the fact that all interactions in physics are spatially local\endnote{This locality is to be understood in the quantum field theory sense, namely that the Hamiltonian of the Standard Model is an integral of a density (and, \emph{a contrario}, a multiple integral in momentum).}. The wavefunction is carved by the Standard Model, all that remains is to paint the branch, and in that restricted sense, the decoherence program has succeeded \cite{schlosshauer2005}. In fact, I am also quite convinced that having a collapse model built from position operators, while simpler, is not necessarily needed. Since the position carving is anyway done by other interactions, the collapse process need only be able to paint (and thus distinguish) two macroscopically distinct branches unambiguously. This can certainly be done with essentially any operator.

\section{A few lessons}

\paragraph*{Quantum mechanics is unbearably rigid} -- ~
We start with an innocent non-linear modification of quantum mechanics, but an apparently mild consistency requirement forces us to have non-linearity vanish exactly at the master equation level. The latter equation, in turn, can be reproduced in many different ways, some of which are even linear at the wave-function level, and thus within the realm of standard quantum mechanics. We try to depart from quantum mechanics without having everything go down the drain, and yet what we obtain is reproducible by quantum theory understood broadly.

\paragraph*{Objective randomness is painfully indistinguishable from effective randomness} -- ~
The second related lesson, which is likely less surprising, is that unequivocally attributing the irreducible unpredictability of quantum mechanics to objective randomness is difficult if not impossible. Naturally, Bohmian mechanics is deterministic and allows to derive the quantum formalism, so we know already that intrinsic randomness is not necessary to explain the irreducible unpredictability of standard quantum mechanics. But we could have expected something weaker: that there exist small modifications of quantum mechanics where intrinsic randomness can be unequivocally pinned down. This is not the case, or at least not in the collapse model way. The randomness collapse models add to quantum mechanics is empirically the same as the one that is already there.

\paragraph*{The split between ontology and empirical content is subtle} -- ~ It is often thought that the choice of ontology of a collapse model (\eg $\text{GRW}_0$, $\text{GRW}_m$, or $\text{GRW}_f$) is merely an aesthetic problem, while the random jump evolution is something more empirical and objective. Actually, the latter is just as metaphysical (I do not use the word in a negative sense). The empirical content of collapse models is entirely contained in their master equation. We choose the standard GRW model instead of the cousins I presented for good enough reasons: it does something for us, namely it solves the measurement problem. This ontological clarity is, in my opinion, the main and the most remarkable achievement of the collapse program. It shows that a clear picture of reality is in principle possible, and provides a powerful tool for further theory building\endnote{In particular, it opens the way to the rather transparent construction of semiclassical theories \cite{tilloy2016,tilloy2019}, which although still embeddable in quantum theory as they ultimately yield a linear master equation, would be hard to guess without starting from the collapse process.}. But we do not get more: even if collapse models are empirically correct, the existence of quantum jumps remains painfully undecidable.

\vskip1cm
\noindent \textit{I am indebted to Dustin Lazarovici, Nicolas Gisin, and \.{I}zel Sar{\i} for their sharp comments on this manuscript and encouragement. Initially written for the 2020 FQXI essay contest.}

\printendnotes

\pagebreak
\bibliographystyle{apsrev4-2}
\bibliography{main}

\end{document}